VRAIN
UPV

# INTELIGENCIA ARTIFICIAL Y EMPLEO PERSPECTIVA TERRITORIAL Y DE GÉNERO

ANTONI MESTRE GASCÓN
XAVIER NAYA PONS
MANOLI ALBERT ALBIOL
VICENTE PELECHANO FERRAGUD

Trabajo financiado por el Ministerio de Ciencia e Innovación bajo el proyecto PRODIGIOUS PID2023-146224OB-I00

Contacto: anmesgas@upv.es

# Tabla de contenido

# 1. Introducción

La acelerada expansión de la inteligencia artificial (IA) en los últimos años ha reconfigurado de manera sustancial los debates sobre el futuro del trabajo, la estructura productiva y la cohesión social. Desde la irrupción de los modelos de aprendizaje profundo y, especialmente, de los modelos fundacionales y generativos, la IA ha pasado de ser una tecnología complementaria a ocupar un papel central en procesos productivos, administrativos y de servicios. Esta transformación ha provocado un creciente interés por parte de instituciones públicas, organizaciones sindicales, centros de investigación y organismos internacionales, que coinciden en señalar que el impacto de la IA en el empleo no solo será significativo, sino también heterogéneo en términos sectoriales, territoriales y de género.

En este contexto, España constituye un caso de estudio especialmente revelador debido a su estructura económica diversificada, con una fuerte especialización en servicios, un peso considerable de actividades turísticas y comerciales, y un tejido industrial en transformación. Las diferencias provinciales y autonómicas en composición sectorial, demografía laboral y nivel de digitalización sugieren que el impacto de la inteligencia artificial será desigual, pudiendo amplificar asimetrías existentes o generar nuevas brechas territoriales. A su vez, estas dinámicas no afectarán por igual a hombres y mujeres: la literatura reciente alerta de una posible intensificación de sesgos estructurales en el mercado laboral España, particularmente en sectores feminizados, en tareas administrativas rutinarias y en ocupaciones de atención al cliente.

Frente a este escenario, disponer de herramientas objetivas y comparables para medir la exposición del empleo a la IA es una necesidad urgente. Sin embargo, la investigación todavía se encuentra en una fase relativamente temprana: los estudios más influyentes —como *AI Occupational Exposure* del NBER, el *Working with AI* del OpenAI & OpenResearch, o los análisis prospectivos de la OCDE— ofrecen metodologías robustas pero centradas en clasificaciones ocupacionales (SOC/ESCO), no sectoriales (CNAE). Esto supone una limitación importante para los análisis territoriales en España, donde la desagregación estadística principal es la actividad económica (CNAE) por provincia y comunidad autónoma. Así, la brecha metodológica entre lo que cuantifican los estudios internacionales y las unidades estadísticas disponibles en España dificulta la traslación directa de sus resultados a políticas públicas de empleo, prospectiva laboral o estrategias sindicales.

El presente informe preliminar se propone precisamente cubrir ese vacío, adaptando las métricas de exposición a la IA procedentes de la literatura internacional al contexto español mediante la construcción de una matriz de incidencia IA–CNAE, diseñada a partir de la evidencia comparada. Esta matriz permite asignar a cada actividad económica un “factor de incidencia” basado en la combinación ponderada de los grupos ocupacionales que típicamente componen cada sector productivo. Con ello, y empleando los microdatos provinciales del Instituto Nacional de Estadística (INE) para 2021–2023, se obtiene una aproximación cuantitativa al grado de aplicabilidad de la IA sobre el empleo en cada provincia y comunidad autónoma, separando además los resultados por sexo.

Es imprescindible subrayar que se trata de un estudio preliminar cuyo objetivo principal es generar una metodología replicable, transparente y abierta, no ofrecer predicciones cerradas. Por ello, los resultados deben interpretarse como una estimación inicial, útil para orientar agendas de investigación, planificación estratégica y toma de decisiones informadas, pero en ningún caso como valores definitivos. La propia naturaleza cambiante de la IA —cuyas capacidades crecen de forma exponencial en ciclos de meses, no de años— exige precaución, flexibilidad y revisión constante.

Aun así, este análisis ofrece varias aportaciones de relevancia para los actores sociales:

1. **Proporciona una visión territorial comparada**, mostrando qué provincias y comunidades autónomas presentan mayor exposición relativa a la IA en función de su composición sectorial.
2. **Permite estudiar las diferencias de género** en la exposición tecnológica, poniendo de manifiesto patrones de riesgo diferencial que pueden requerir políticas específicas.
3. **Construye una herramienta metodológica trasladable** a futuras ediciones del Censo de Población Ocupada, permitiendo comparar tendencias año a año.
4. **Facilita el diálogo entre instituciones, organizaciones sindicales y agentes económicos**, proporcionando un lenguaje común para discutir la transición tecnológica.

En un momento en el que España está desplegando estrategias nacionales de digitalización e inteligencia artificial —como la Estrategia Nacional de IA (ENIA), el PERTE de la Nueva Economía de la Lengua o los planes de digitalización industrial— contar con un análisis inclusivo, territorialmente desagregado y sensible al género no es solo recomendable, sino necesario para asegurar una transición justa. La IA tiene el potencial de aumentar la productividad, mejorar servicios públicos y generar oportunidades de empleo cualificado; pero también puede intensificar desigualdades si no se anticipan sus efectos de manera integral.

Por ello, este informe se sitúa deliberadamente en la intersección entre investigación empírica y diagnóstico estratégico, buscando no solo ofrecer datos, sino también orientar las preguntas que deberán guiar el debate en los próximos años:

- ¿En qué territorios se concentra mayor riesgo tecnológico y por qué?
- ¿Existen brechas de género en la exposición a la IA que deban ser abordadas con urgencia?
- ¿Qué sectores deberían ser prioritarios en estrategias de recualificación laboral?
- ¿Qué papel deben desempeñar las organizaciones sindicales y las administraciones públicas en la gobernanza de esta transición?

El resto del documento desarrolla estas cuestiones a través de una revisión del estado del arte, una descripción detallada de la metodología empleada para construir la matriz de incidencia IA–CNAE, la presentación de los resultados territoriales y de género para el periodo 2021–2023 y, finalmente, unas conclusiones orientadas a la acción.

# 2. Estado del Arte

En los últimos años se ha consolidado una literatura amplia y heterogénea sobre el impacto de la inteligencia artificial (IA) en el trabajo, con cuatro líneas principales: (1) la medición de la exposición ocupacional a la IA, (2) el análisis específico de la IA generativa, (3) los impactos diferenciales por género, y (4) la dimensión territorial. Este informe se apoya en estos cuatro pilares para construir una matriz de incidencia CNAE-IA aplicada al territorio y al género.

## 2.1. De la automatización clásica a la IA generativa

La literatura clásica sobre tecnología y empleo se articuló en torno al análisis de tareas rutinarias y su susceptibilidad a la automatización. Los trabajos seminales de Autor, Levy y Murnane [1] impulsaron el enfoque *task-based*, que desplaza la atención desde las ocupaciones completas hacia las tareas específicas que las componen. Con la irrupción reciente de la IA moderna, y en particular de los modelos fundacionales y la IA generativa, el perímetro de lo técnicamente automatizable se amplía considerablemente: tareas no rutinarias, de naturaleza cognitiva o lingüística, pasan a ser abordables por sistemas avanzados de IA [2], [3].

La consecuencia es un cambio conceptual profundo: ya no solo interesa qué tareas pueden automatizarse, sino con qué grado de complementariedad o sustitución se relacionan con sistemas inteligentes capaces de interpretar, sintetizar o generar información.

## 2.2. Medición de la exposición ocupacional a la IA

Los estudios recientes sobre exposición ocupacional a la IA comparten un enfoque metodológico común: (1) caracterizar las capacidades de la IA, (2) relacionarlas con descriptores de tareas u ocupaciones, y (3) construir índices de exposición o aplicabilidad extrapolables a sectores y territorios.
El presente informe se apoya de forma central en *Working with AI: Measuring the Applicability of Generative AI to Occupations*, de Tomlinson et al. [4], que utiliza más de 200.000 conversaciones laborales asistidas por Microsoft Copilot/Bing. Este trabajo identifica qué actividades laborales ya se realizan con IA, con qué éxito y en qué amplitud de ocupaciones, derivando un AI applicability score que muestra niveles especialmente altos en ocupaciones de informática, administración, ventas y finanzas.

Este enfoque se complementa con estudios como *GPTs are GPTs* [5], que estima que alrededor del 80 % de los empleos en Estados Unidos tiene al menos un 10 % de tareas potencialmente afectadas por los modelos de lenguaje, y con los análisis europeos de Nurski [6], que adaptan estos índices al contexto de la UE considerando simultáneamente potencial de sustitución y complementariedad. Nuestro trabajo se inscribe en esta tradición: partimos de medidas ocupacionales y las trasladamos a la clasificación sectorial CNAE para posibilitar un análisis territorial y de género.

## 2.3. IA generativa, productividad y desempeño laboral

Más allá de la exposición teórica, emergen evidencias empíricas sobre los efectos de la IA generativa en el desempeño laboral. Experimentos de campo recientes muestran que la introducción de asistentes generativos mejora la productividad y la calidad en tareas de redacción, servicio al cliente o programación, con incrementos particularmente intensos en trabajadores menos experimentados [7].
En el plano macroeconómico, estudios del Banco de España [8] y del Banco Central Europeo [9] sugieren que aproximadamente un 10 % del tiempo de trabajo podría transformarse o acelerarse de manera sustancial en horizontes relativamente cortos.

La literatura converge en que, en su fase inicial, la IA generativa actúa como complemento cognitivo más que como sustituto pleno, especialmente en tareas que requieren juicio contextual, supervisión humana o responsabilidad legal. Sin embargo, a medio plazo pueden producirse reconfiguraciones ocupacionales, cambios en la demanda de habilidades y efectos distributivos que dependen de la estructura productiva de cada territorio.

## 2.4. Dimensión territorial y de género

La distribución territorial de los efectos de la IA es desigual. Estudios recientes muestran que regiones con mayor especialización en sectores intensivos en IA tienden a beneficiarse más de la adopción tecnológica, mientras que otras regiones pueden enfrentar riesgos superiores de sustitución [10]. Los informes de la OCDE sobre empleo local [11] subrayan que las áreas metropolitanas —con mayor concentración de empleo cualificado— absorben antes las oportunidades derivadas de la IA, a diferencia de regiones periféricas dependientes de actividades de menor productividad.

En España, la evidencia apunta a una fuerte concentración de empresas y empleo ligado a IA en Madrid y Cataluña, con nodos relevantes en la Comunitat Valenciana, Andalucía y País Vasco [12]. Esto sugiere que los beneficios económicos derivados de la IA podrían distribuirse territorialmente de forma desigual incluso en contextos donde la exposición técnica a la automatización es más homogénea.

El impacto de la IA por género es un elemento central del debate actual. El informe *Algorithm and Eve* de la OCDE [13] muestra que la exposición ocupacional a la IA es, en promedio, similar entre mujeres y hombres, pero advierte que las mujeres están sobrerrepresentadas en ocupaciones administrativas altamente automatizables e infrarrepresentadas en ocupaciones STEM con mayor potencial de complementariedad con la IA. Un estudio reciente del BCE [14] encuentra que, en países con elevada participación y nivel educativo femenino, la difusión de la IA se asocia con un aumento de la participación laboral de las mujeres precisamente en ocupaciones más expuestas.

Por otro lado, la OIT alerta de que las mujeres son entre dos y tres veces más propensas a trabajar en ocupaciones altamente automatizables por IA generativa, especialmente en tareas clericales [15]. En España, la evidencia indica que las mujeres tienden a concentrarse en titulaciones universitarias con menor exposición a IA y menores retornos salariales [16].

El enfoque de este informe introduce una novedad: combinamos la estructura sectorial (CNAE), la distribución territorial y la matriz de aplicabilidad IA-CNAE para estimar cómo la composición productiva local amplifica o mitiga el impacto diferencial por género.

## 2.6. Vacíos de la literatura y aportación del informe

Pese a los avances recientes, persisten tres vacíos relevantes:

(1) la mayoría de índices de exposición están formulados a nivel ocupacional, sin matrices consensuadas por CNAE;

(2) los análisis territoriales suelen centrarse en niveles NUTS2/NUTS3, sin alcanzar el nivel provincial español;

(3) la literatura de género presta menos atención a la interacción entre género, territorio y estructura sectorial.

Este informe aborda estos vacíos mediante la construcción de una matriz IA-CNAE basada en evidencia ocupacional consolidada [4], aplicada después a los datos de empleo por provincia, comunidad autónoma y sexo, generando una primera cartografía territorial y de género de la aplicabilidad de la IA en el mercado laboral español.

# 3. Metodología

La metodología empleada en este informe nace de una necesidad evidente: mientras la mayor parte de la literatura internacional sobre inteligencia artificial se formula en términos ocupacionales —habitualmente a partir de las clasificaciones SOC, ISCO o ESCO—, la estadística oficial española ofrece una fotografía del mercado laboral basada fundamentalmente en sectores (CNAE) y no en ocupaciones. Para poder analizar el impacto territorial y de género de la IA con el nivel de detalle que requiere la realidad española, es imprescindible construir un puente entre ambos mundos conceptuales. Este informe, de carácter preliminar, parte precisamente de esa motivación metodológica.

El primer paso del proceso consiste en seleccionar y adaptar un marco conceptual de exposición a la IA que sea compatible con nuestra unidad de análisis. El estado del arte ofrece diversas aproximaciones, pero uno de los estudios más influyentes es el índice de *AI Applicability* formulado por Tomlinson et al. en su estudio *Working with AI* [4]. A diferencia de trabajos anteriores centrados en la automatización clásica, esta propuesta no mide únicamente la sustituibilidad de tareas, sino su grado de aplicabilidad para la IA generativa: hasta qué punto la tecnología puede colaborar, asistir o intervenir en el desempeño laboral real. El índice se construye a partir de evidencia empírica robusta (más de 200.000 interacciones laborales con IA), lo que permite identificar con precisión en qué tipo de tareas la IA generativa es ya capaz de aportar valor.

Este enfoque proporciona un marco ocupacional muy rico, pero es necesario traducirlo a la estructura sectorial española. A tal efecto, se llevó a cabo un proceso de correspondencias conceptuales entre los grandes grupos ocupacionales SOC y las ramas de actividad económica CNAE. Para cada CNAE a dos dígitos, se analizaron sus perfiles ocupacionales predominantes, las tareas características del sector, su grado de digitalización y la evidencia disponible sobre adopción de tecnologías de IA. Este ejercicio de traducción no busca una equivalencia perfecta —inexistente dada la heterogeneidad interna de muchos sectores—, sino una representación razonable de cuáles son las capacidades de la IA que resultan relevantes en cada rama de actividad.

A partir de esta correspondencia se construyó una matriz sectorial CNAE–IA (Anexo II) que asigna a cada sector un factor de incidencia comprendido entre 0,06 y 0,30. Sectores con predominio de tareas manuales, físicas o rutinarias (como la agricultura, la pesca o las industrias extractivas) reciben factores bajos; sectores caracterizados por procesos administrativos, trabajo lingüístico, actividades de síntesis o alto contenido analítico (comercio, información y comunicaciones, actividades profesionales, finanzas) reciben factores medios o altos; y aquellos con un peso significativo de perfiles técnicos avanzados (informática, telecomunicaciones, ingeniería) alcanzan los niveles más elevados. La matriz no es un reflejo mecánico de ninguna clasificación única, sino la síntesis ponderada de la literatura disponible, de analogías entre grupos ocupacionales y de consideraciones sectoriales propias del caso español.

Para el análisis empírico se emplearon los microdatos del Censo anual de población ocupada del INE (series 69960-), correspondientes a los años 2021 y 2022, y desagregados por provincia, sexo y sector CNAE a dos dígitos. Esta fuente permite reconstruir la estructura productiva de cada territorio en términos de empleo sectorial, proporcionando así la base estadística necesaria para aplicar la matriz IA–CNAE. El procedimiento de cálculo es directo: para cada combinación provincia–sector se multiplica el empleo registrado en ese sector por el factor de incidencia asignado al CNAE correspondiente; la suma de estos productos proporciona una estimación del "empleo equivalente" potencialmente afectado por la IA en la provincia, mientras que el cociente entre este valor y el empleo total provincial ofrece un indicador sintético de exposición relativa a la IA. El mismo proceso se replica por sexo, lo que permite identificar diferencias estructurales entre el empleo femenino y masculino.

La aplicación de esta metodología al conjunto del territorio español ofrece una cartografía coherente con el conocimiento cualitativo disponible sobre la estructura económica del país. Las provincias con mayor peso de servicios avanzados —como Madrid, Barcelona o Málaga— aparecen con valores más elevados de exposición, mientras que aquellas con un peso mayor de sectores primarios o industriales tradicionales muestran niveles relativamente menores. Las diferencias de género también reproducen patrones ampliamente documentados: la especialización femenina en tareas administrativas, educativas, sanitarias y de servicios sociales, junto con su menor presencia en actividades STEM, incrementa ligeramente la exposición estimada para las mujeres en prácticamente todos los territorios.

Es fundamental subrayar que el modelo aquí empleado proporciona una aproximación preliminar**,** no una estimación definitiva. La traslación desde ocupaciones SOC a sectores CNAE implica necesariamente simplificaciones: no todas las empresas dentro de un mismo sector presentan la misma composición ocupacional, la misma intensidad tecnológica ni la misma velocidad de adopción de IA. Tampoco se abordan, en esta fase, efectos dinámicos como la creación de nuevas tareas, la reorganización del trabajo o los cambios en la productividad inducidos por la IA. El modelo mide exposición potencial**,** no sustitución real, y por ello debe interpretarse como una brújula orientativa más que como un pronóstico detallado.

Con todo, esta metodología constituye una base sólida y coherente para comenzar a evaluar la interacción entre tecnología, empleo y territorio en España. Su principal valor reside en permitir que los datos sectoriales del INE —ampliamente difundidos y comparables en el tiempo— puedan incorporarse al debate internacional sobre el impacto de la IA, sin perder de vista las especificidades territoriales y las desigualdades de género. En futuras versiones del informe, este enfoque podrá refinarse incorporando datos ocupacionales más desagregados, análisis de tareas directamente vinculados a modelos como GPT-4 o Gemini, y estimaciones contrafactuales sobre efectos de desplazamiento y complementariedad.

## 3.1. Modelo de cálculo

Una vez definida la matriz de incidencia CNAE–IA, el siguiente paso consiste en proyectarla sobre la estructura sectorial real de cada territorio. Para ello se ha desarrollado un modelo de cálculo sencillo pero robusto, que permite estimar la magnitud relativa del empleo potencialmente afectado por la inteligencia artificial en cada provincia y comunidad autónoma. Este modelo se basa en el principio de proporcionalidad: se asume que la exposición total de un territorio es la suma de las exposiciones de cada uno de sus sectores productivos, ponderada por el tamaño del empleo en cada rama de actividad.

El procedimiento consta de dos niveles de agregación. En primer término, se calcula la exposición sectorial dentro de cada provincia; posteriormente, se integran estos valores para obtener un indicador territorial sintético. La lógica del cálculo es lineal y se expresa mediante las siguientes formulaciones matemáticas.

### 3.1.1. Exposición IA por sector en cada provincia

Para cada provincia *p* y sector CNAE *c*, se define la magnitud:

$$IA_\text{empleo}(p,c) = empleo(p,c) \times factorIA(c)$$

donde:

- *empleo(p,c)* representa el número de personas ocupadas en la provincia *p* en el sector *c*, según los microdatos del INE;
- *factorIA(c)* es el coeficiente de incidencia de la IA asignado al sector *c* en la matriz CNAE–IA.

El término *IA_empleo(p,c)* puede interpretarse como un empleo equivalente ajustado por exposición: no es un número de puestos concretos, sino una magnitud sintética que cuantifica cuánta parte del empleo provincial se encuentra en sectores con mayor aplicabilidad potencial de la IA. Sectores con valores altos del factor IA contribuyen de forma más relevante al total provincial, y viceversa.

Este enfoque permite trasladar la incidencia relativa de cada sector a un único indicador agregable.

### 3.5.2. Exposición IA agregada por provincia y comunidad autónoma

Una vez obtenidos los valores *IA_empleo(p,c)* para todos los sectores de una provincia, es posible construir un indicador sintético de exposición territorial. Este se define como:

$$IA_\text{share}(p) = \frac{\sum_c IA_\text{empleo}(p,c)}{\sum_c empleo(p,c)}$$

El numerador representa la suma del empleo ajustado por exposición para todos los sectores de la provincia; el denominador, el empleo total provincial. El cociente resultante es un indicador en escala 0–1 que expresa la proporción del empleo provincial cuya estructura sectorial presenta una alta aplicabilidad potencial de la IA**.**

Este indicador puede interpretarse como una medida comparativa entre territorios: no indica sustitución tecnológica real, sino grado relativo de exposición estructural, lo que permite identificar provincias con un tejido productivo más vulnerable o más intensivo en tareas susceptibles de transformarse mediante IA.

Mediante un procedimiento idéntico se obtienen valores agregados a nivel de comunidad autónoma, reemplazando la provincia *p* por la comunidad autónoma *a*:

$$IA_\text{share}(a) = \frac{\sum_{p\in a}\sum_c IA_\text{empleo}(p,c)}{\sum_{p\in a}\sum_c empleo(p,c)}$$

Esta extensión permite comparar territorios en distintas escalas administrativas.

### 3.5.3. Exposición por género

La metodología permite también capturar las diferencias de exposición entre hombres y mujeres. Para ello, simplemente se sustituye el dato de empleo total provincial por el empleo desagregado por sexo:

- $empleo^{M}(p,c)$: empleo femenino en la provincia *p* y sector *c*;
- $empleo^{H}(p,c)$: empleo masculino en la provincia *p* y sector *c*.

Las expresiones quedan así:

$$IA_\text{share}^M(p) = \frac{\sum_c empleo^M(p,c) \times factorIA(c)}{\sum_c empleo^M(p,c)}$$

$$IA_\text{share}^H(p) = \frac{\sum_c empleo^H(p,c) \times factorIA(c)}{\sum_c empleo^H(p,c)}$$

Estos indicadores permiten analizar la brecha tecnológica de género desde una perspectiva estructural: no en función de tareas individuales o cualificaciones, sino de la diferente composición sectorial del empleo masculino y femenino. La comparación entre ambos valores revela hasta qué punto la estructura productiva de cada territorio sitúa a mujeres y hombres en posiciones diferenciales frente a la difusión de la IA.

## 3.2. Limitaciones del modelo

Dado el carácter preliminar de este informe, es necesario explicitar las limitaciones metodológicas que acompañan al ejercicio realizado. El enfoque desarrollado proporciona una aproximación coherente y útil para comprender la exposición territorial y de género a la inteligencia artificial, pero debe interpretarse con cautela y no como una medición exhaustiva o definitiva.

En primer lugar, la metodología trabaja a partir de una desagregación sectorial, no de tareas ni de ocupaciones. El uso del CNAE como unidad de análisis implica asumir que cada sector tiene una estructura interna relativamente homogénea en términos de composición de tareas. Sin embargo, dentro de un mismo sector conviven empresas con realidades tecnológicas muy dispares, así como ocupaciones cuya exposición a la IA puede diferir de forma sustancial. La ausencia de una capa de desagregación por tareas limita, inevitablemente, la precisión con la que puede estimarse la incidencia real de la IA dentro de cada rama de actividad.

En segundo lugar, el modelo asume una composición ocupacional promedio dentro de cada sector, basada en análisis previos y en patrones generales derivados de la literatura. La falta de datos ocupacionales sistemáticos por provincia impide verificar hasta qué punto la estructura real de un sector concreto en un territorio determinado coincide con la composición estimada. Este paso es metodológicamente necesario para permitir comparaciones territoriales, pero introduce un nivel de abstracción que debe ser reconocido.

Una tercera limitación deriva del carácter estático del modelo. La aproximación utilizada mide la exposición potencial en un momento dado, pero no captura efectos dinámicos asociados a la difusión tecnológica, tales como la creación de nuevas tareas, la reasignación de funciones dentro de una empresa, la transformación de ocupaciones existentes o la emergencia de puestos de trabajo vinculados a la supervisión, implementación o mejora de sistemas de IA. La literatura muestra que este tipo de efectos pueden ser significativos y, en algunos casos, compensar parcialmente los riesgos de automatización.

Asimismo, la matriz CNAE–IA debe entenderse como una aproximación conceptual, no como un valor empírico directo. Su construcción combina la evidencia disponible sobre exposición ocupacional a la IA generativa con decisiones analíticas inevitables al trasladar dicho conocimiento a una clasificación sectorial. A medida que se desarrollen nuevos estudios, especialmente aquellos que incorporen tareas específicas vinculadas a modelos fundacionales o a la automatización cognitiva, será posible refinar esta matriz y ajustar los factores de incidencia asignados a cada sector.

Por último, conviene subrayar que el modelo evalúa exposición potencial, no sustitución real de empleo. El hecho de que un territorio presente un valor elevado de IA-share no implica necesariamente que se produzcan pérdidas de empleo en esa magnitud. La literatura internacional señala que la IA generativa tiene un componente significativo de complementariedad, especialmente en actividades cognitivas y lingüísticas, y que sus efectos netos sobre el empleo dependen de múltiples factores: capacidad de absorción tecnológica, inversión en digitalización, nivel educativo de la fuerza laboral, políticas de formación y mecanismos de transición justa.

Estas limitaciones no invalidan los resultados del informe. Al contrario, permiten contextualizarlos adecuadamente. El modelo proporciona una visión estructural y comparativa del grado de exposición de los territorios españoles a la inteligencia artificial, útil para la planificación estratégica, la elaboración de diagnósticos y el diseño de políticas públicas o sindicales. Sin embargo, sus conclusiones deben leerse como estimaciones preliminares, sujetas a revisión y mejora conforme avance el conocimiento científico y se disponga de datos más detallados.

# 4. Resultados

La aplicación de la matriz CNAE–IA a los datos provinciales y autonómicos de empleo permite identificar una serie de patrones territoriales y de género que estructuran el impacto potencial de la inteligencia artificial en el mercado laboral español. Aunque este es un ejercicio preliminar —y así debe interpretarse—, los resultados muestran una consistencia notable entre años, territorios y grupos poblacionales, lo que sugiere que la exposición a la IA responde más a elementos estructurales del tejido productivo que a fluctuaciones coyunturales del ciclo económico.

## 4.1. Exposición agregada de España

El primer resultado relevante es que la estructura económica española presenta una exposición a la IA moderadamente alta, con valores que oscilan entre el 18% y el 22% del empleo total**,** dependiendo de la provincia y del año. Esta horquilla estrecha indica que la IA posee un potencial transversal de transformación sobre el conjunto del mercado laboral, incluso en sectores tradicionalmente considerados menos sustituibles. Sin embargo, la magnitud exacta de esa exposición varía sustancialmente en función del peso relativo de los servicios avanzados, la administración, las actividades comerciales y los sectores de información y comunicaciones en cada territorio.

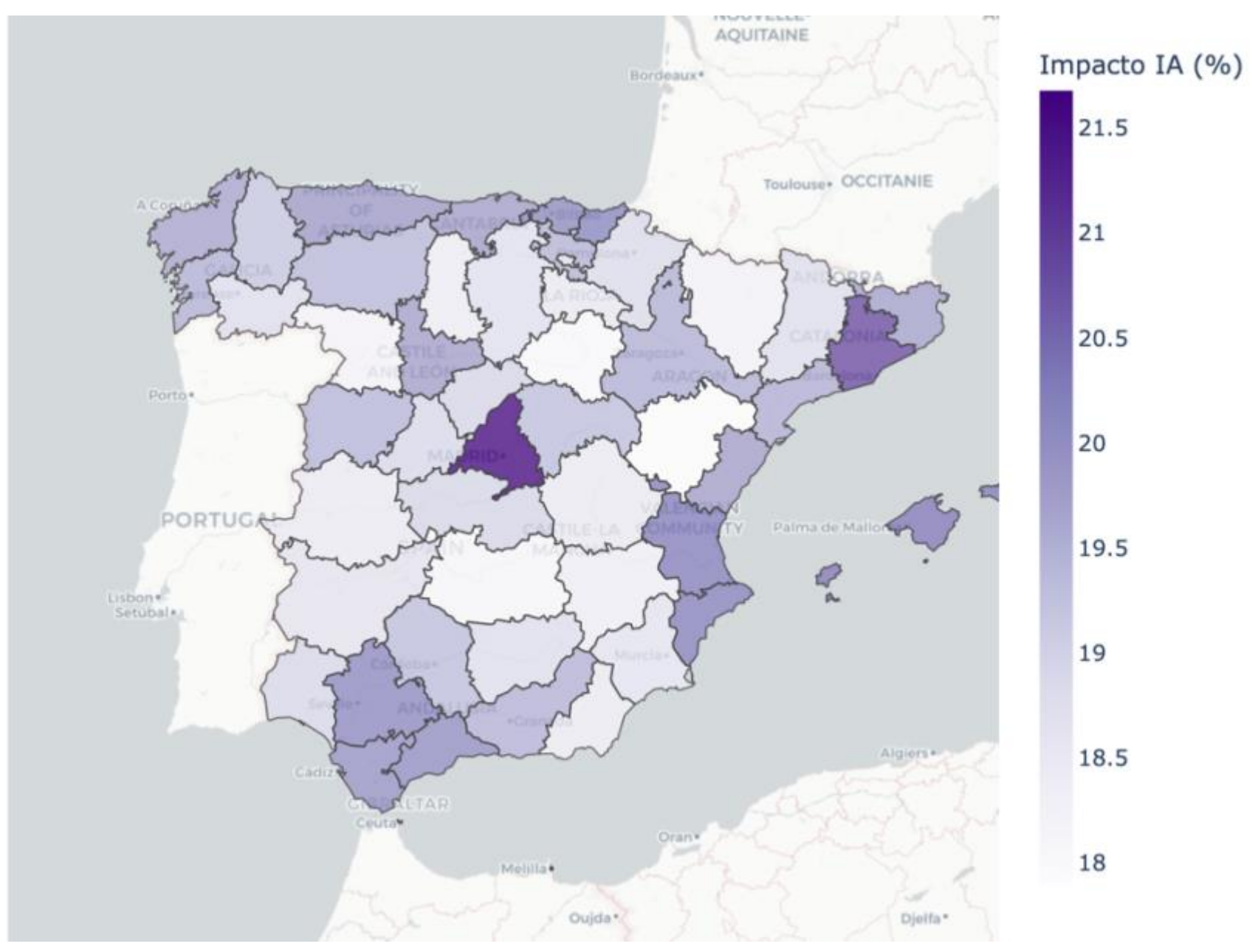

*Figura 1. Incidencia de la IA en el empleo, datos del 2022.*

El empleo femenino se sitúa sistemáticamente en el tramo superior de exposición, mientras que el empleo masculino se concentra en la parte inferior. Esta diferencia estructural, que analizaremos más adelante, es coherente con la evidencia internacional sobre la mayor vulnerabilidad de tareas administrativas, educativas, sanitarias o de servicios personales —todas ellas intensamente feminizadas—.

## 4.2. Distribución territorial de la exposición (2021–2022)

Uno de los patrones más nítidos del análisis es la clara polarización territorial en el IA-share. Existe un corredor centro-mediterráneo que agrupa a las provincias con mayores niveles de exposición: Madrid, Barcelona, Valencia, Alicante y Málaga, junto con las dos capitales insulares, Las Palmas y Santa Cruz de Tenerife.

Estas provincias superan de forma consistente el 20 % de exposición, llegando Madrid a valores superiores al 21,5 % en ambos años.

Este patrón es coherente con la especialización funcional de estos territorios. Las áreas metropolitanas albergan una mayor proporción de actividades financieras, informáticas, de servicios empresariales, actividades profesionales y administración corporativa, todas ellas intensivas en tareas cognitivas y lingüísticas que presentan alta aplicabilidad de IA generativa. También concentran un porcentaje elevado de empleo en comercio, educación y sanidad, sectores que la literatura identifica como particularmente susceptibles de incorporar IA en tareas de gestión, atención al cliente, planificación, documentación o síntesis informativa.

En contraste, provincias del interior peninsular y con una fuerte implantación de sectores primarios o manufacturas tradicionales presentan niveles notablemente inferiores de exposición. Soria, Zamora, Teruel, Cuenca o Palencia aparecen de forma consistente en el tramo bajo de la distribución, con valores próximos al 17,5–18,5 %. En términos macroterritoriales, Castilla y León, Castilla-La Mancha y Aragón muestran una exposición estructural inferior a la media nacional.

Es relevante observar que estas diferencias territoriales no se han modificado sustancialmente entre 2021 y 2022. La clasificación relativa de las provincias permanece extraordinariamente estable, con variaciones que rara vez superan las dos décimas. Esto refuerza la idea de que la exposición a la IA está determinada ante todo por la estructura productiva estable de cada territorio, y no por oscilaciones puntuales en el empleo sectorial.

## 4.3. Exposición por género: una brecha estructural y persistente

La dimensión de género introduce un eje adicional de interpretación que enriquece el análisis territorial. En todas las provincias, sin excepción, el empleo femenino presenta mayor exposición a la IA que el empleo masculino. Esta brecha oscila mayoritariamente entre 1,5 y 3 puntos porcentuales, aunque en territorios altamente terciarizados puede superar ese margen.

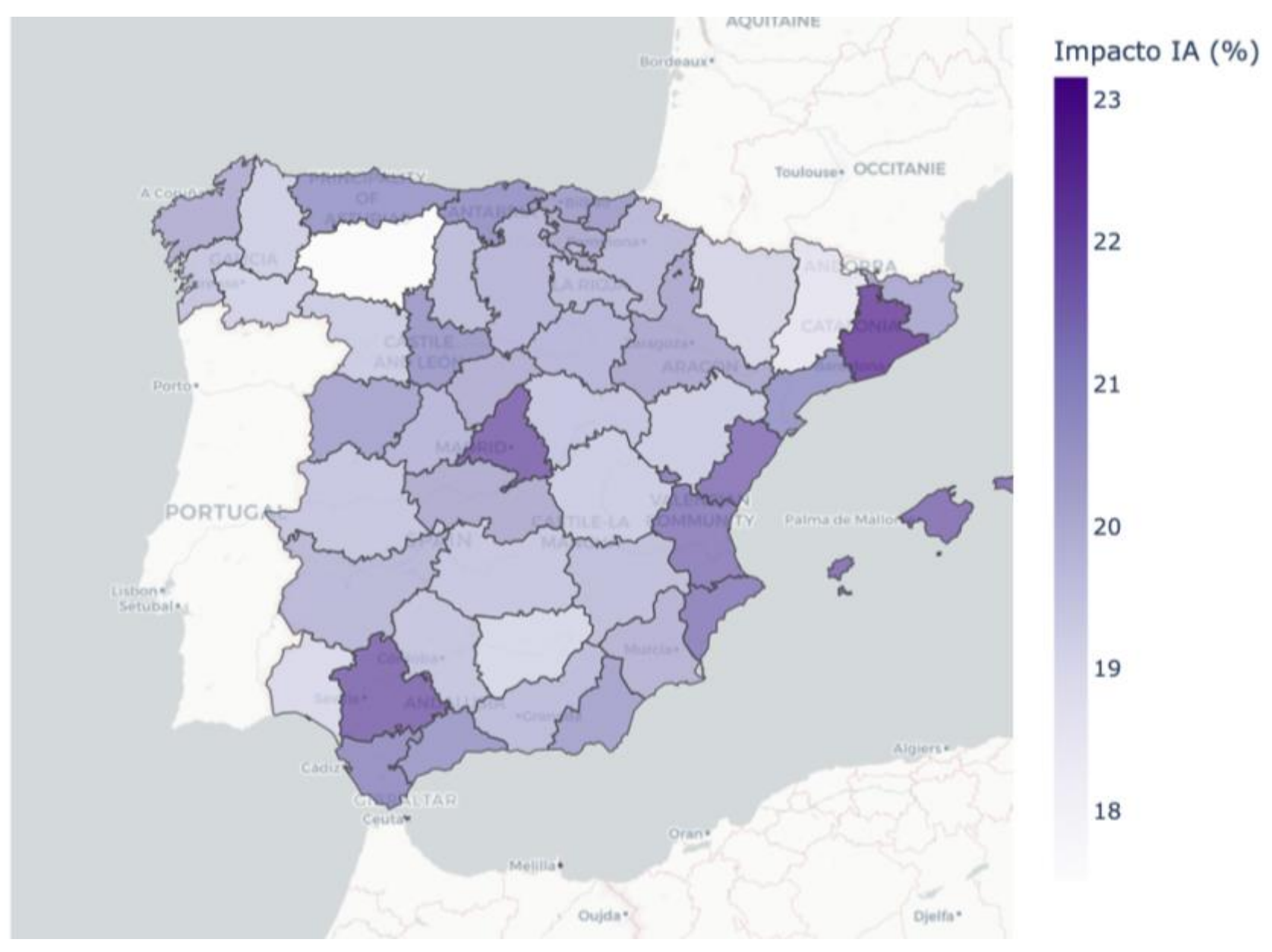

*Figura 2. Incidencia de la IA en el empleo de mujeres, datos del 2022.*

En el caso del empleo femenino, los valores más elevados se registran en Madrid, Illes Balears, Santa Cruz de Tenerife, Barcelona, Cantabria y Málaga, con exposiciones que alcanzan o superan el 21,5 %. El patrón geográfico es análogo al observado en los totales, pero la intensidad es aún mayor. La razón es doble: por un lado, las mujeres se concentran en sectores con alta aplicabilidad de IA, como educación, sanidad, servicios administrativos, comercio o actividades sociales; por otro, están infrarrepresentadas en sectores de baja exposición, como construcción, transporte o industrias extractivas.

En el caso masculino, aunque persisten los patrones metropolitanos de exposición elevada, los valores son sensiblemente inferiores. Las provincias con mayor exposición masculina —Madrid, Las Palmas, Barcelona o Santa Cruz de Tenerife— se sitúan en torno al 20–21 %, mientras que provincias del interior bajan al 16,5–18 %. Dado que los hombres presentan una mayor concentración en industria y construcción, su exposición relativa es menor en términos estructurales.

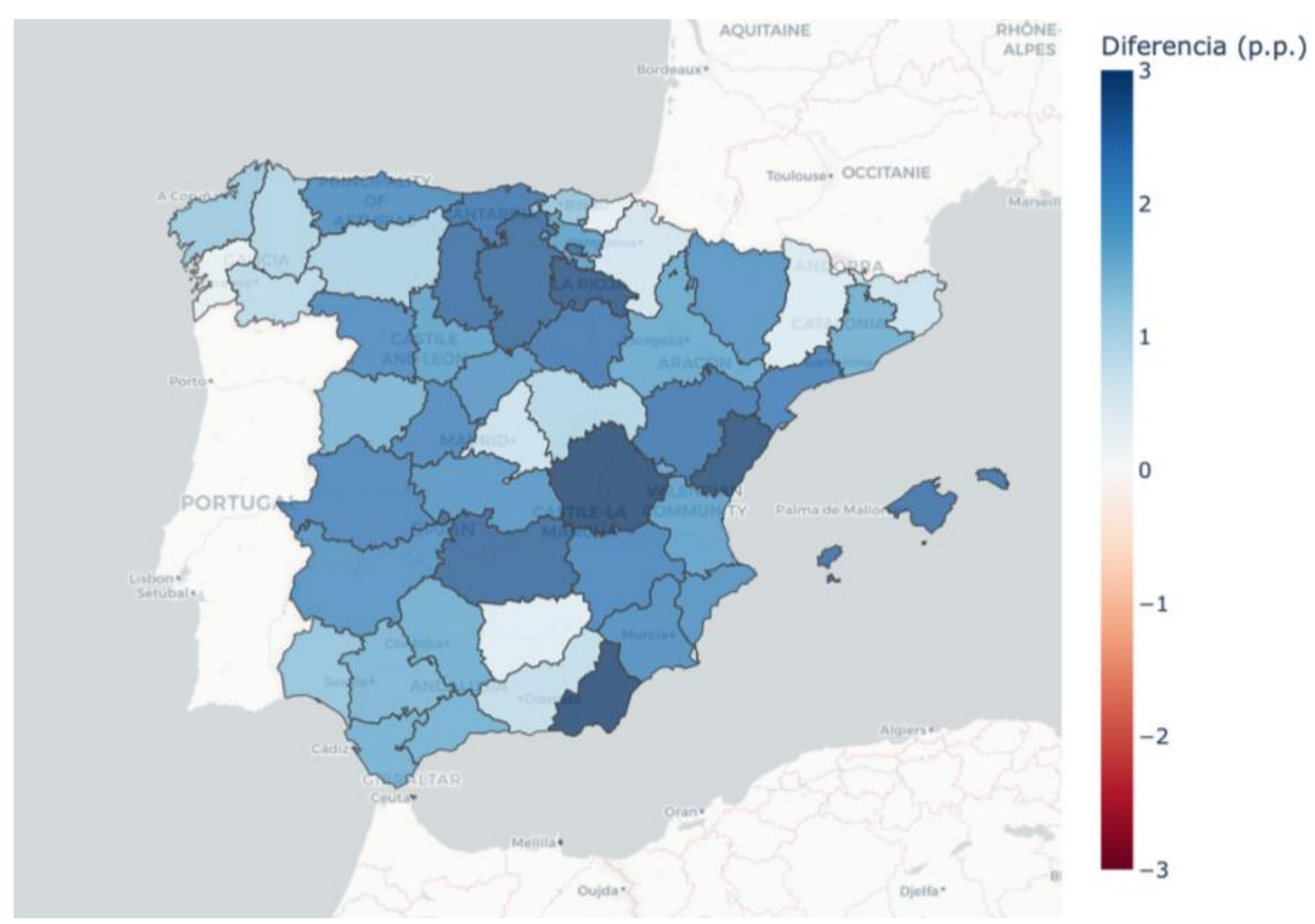

*Figura 3. Diferencia entre la incidencia de la IA entre hombres (rojo) mujeres (azul).*

La brecha de género resultante es un fenómeno de carácter nacional y no local. No existe una sola provincia donde el IA-share masculino supere al femenino, lo que pone de manifiesto una vulnerabilidad estructural diferencial que deberá ser tenida en cuenta en políticas de transición tecnológica, formación y protección social.

## 4.4. Diferencias sectoriales y su reflejo territorial

La matriz CNAE–IA permite profundizar en los factores que explican por qué ciertos territorios presentan mayores niveles de exposición a la inteligencia artificial que otros. Lejos de ser un fenómeno homogéneo, la exposición territorial a la IA refleja un conjunto de especializaciones productivas persistentes, producto de décadas de concentración sectorial, desarrollo económico local y configuración histórica del mercado laboral.

Para comprender estos patrones, es útil examinar qué sectores actúan como motores de exposición y cuáles ejercen un efecto amortiguador, reduciendo la exposición agregada. Esta lectura sectorial ayuda a explicar no solo el nivel de exposición, sino también su distribución geográfica.

Las provincias con mayor IA-share comparten cuatro elementos estructurales.

En primer lugar, presentan una proporción especialmente significativa de empleo en comercio minorista y mayorista, sectores que, según la matriz CNAE–IA, muestran factores de aplicabilidad elevados debido a su fuerte componente de atención al cliente, gestión de inventarios, planificación, interacción lingüística y operaciones administrativas. La IA generativa es especialmente competente en estas tareas, lo que eleva de manera mecánica la exposición territorial en aquellas provincias donde el comercio representa una parte sustancial del tejido económico local.

En segundo lugar, las provincias más expuestas cuentan con un peso elevado de servicios administrativos, educativos y sanitarios, actividades que, aunque diversas en naturaleza, comparten una dependencia estructural de la documentación, la gestión de información, la planificación, la interacción comunicativa y la resolución de problemas de carácter normativo o procedimental. Estos sectores concentran tareas particularmente susceptibles de ser automatizadas —o al menos transformadas— mediante IA generativa, especialmente en funciones de asistencia, redacción, clasificación y síntesis.

Un tercer elemento clave es la presencia destacada de actividades profesionales, científicas y técnicas, un amplio conjunto de servicios de alto valor añadido que engloba consultoría, ingeniería, auditoría, arquitectura, investigación y servicios jurídicos. Aunque estas actividades dependen de competencias cognitivas complejas y de un alto grado de creatividad o responsabilidad, también incluyen una proporción importante de tareas repetitivas, documentales o computacionales que pueden ser asistidas por sistemas avanzados de IA. Su elevada concentración en áreas metropolitanas y capitales provinciales explica, en gran medida, la exposición superior observada en territorios urbanos.

Finalmente, los territorios con mayor exposición suelen albergar una masa crítica de empresas vinculadas a información, telecomunicaciones y servicios financieros, sectores que lideran la adopción y desarrollo de tecnologías digitales. Su presencia actúa como un multiplicador, no solo aumentando la exposición directa del sector, sino contribuyendo a la difusión de prácticas de trabajo intensivas en tecnología hacia otros sectores locales mediante efectos de arrastre, redes de servicios auxiliares y derrames de innovación.

En el extremo opuesto, las provincias con menor exposición comparten una estructura productiva más intensiva en agricultura, manufacturas tradicionales y construcción, sectores que presentan factores IA notablemente inferiores. En la agricultura y la pesca, el carácter marcadamente físico, manual y estacional del trabajo limita, al menos por ahora, la aplicabilidad de modelos generativos o algoritmos cognitivos. Las manufacturas tradicionales —particularmente las asociadas al textil, la madera, el cuero o la metalurgia básica— han experimentado automatizaciones significativas durante décadas, pero muchas de sus tareas siguen siendo manuales, operativas o asociadas al control de maquinaria física, tipos de actividad donde la IA generativa tiene menos capacidad transformadora. La construcción, por su parte, se compone de tareas predominantemente físicas, operativas y altamente contextualizadas, cuya automatización requiere tecnologías robóticas avanzadas más que modelos basados en lenguaje o aprendizaje profundo.

Este gradiente sectorial explica, de manera bastante mecánica, la distribución territorial observada: los territorios urbanos y altamente terciarizados muestran un IA-share elevado, mientras que los territorios rurales o industriales clásicos muestran valores más modestos. Las diferencias no responden, por tanto, a dinámicas coyunturales ni a decisiones recientes de inversión, sino a estructuras económicas estables, profundamente arraigadas y difíciles de modificar en el corto plazo. Esta lectura también arroja luz sobre la notable estabilidad temporal del indicador entre 2021 y 2022.

## 4.5. Evolución temporal y estabilidad estructural

El análisis comparado entre 2021 y 2022 revela una característica central del modelo: la exposición potencial a la inteligencia artificial en España muestra una extraordinaria estabilidad temporal. Las provincias no solo mantienen prácticamente inalterada su clasificación relativa en ambos años, sino que las variaciones absolutas del IA-share son mínimas y rara vez superan unas décimas. Esta estabilidad es coherente con el hecho de que la exposición estimada no mide la adopción real de IA, sino la susceptibilidad estructural del tejido productivo a verse transformado por ella.

En otras palabras, el modelo no captura fluctuaciones coyunturales del ciclo económico ni variaciones anuales en el empleo sectorial, sino la arquitectura profunda de la economía provincial y autonómica. Y esa arquitectura cambia muy lentamente.

Desde una perspectiva metodológica, esta estabilidad temporal funciona como una forma indirecta de validación interna del modelo. Si las variaciones hubieran sido abruptas o incoherentes —por ejemplo, si provincias agrícolas hubieran mostrado incrementos significativos de un año a otro— se habría puesto en cuestión la solidez de la metodología. Por el contrario, la constancia de los valores y la coherencia del patrón de distribución territorial sugieren que el indicador está captando rasgos estructurales y no ruido estadístico.

La literatura económica respalda esta interpretación: las transformaciones sectoriales profundas se desarrollan a lo largo de horizontes prolongados, y los cambios relevantes en la exposición tecnológica provienen más de la reconfiguración del tejido productivo (por ejemplo, desplazamiento hacia sectores de alto valor añadido, crecimiento del sector digital, externalización de servicios profesionales) que de oscilaciones anuales del empleo. De forma similar, la adopción empresarial de tecnologías avanzadas no se produce de manera uniforme ni instantánea, sino a través de procesos graduales de inversión, aprendizaje organizativo y reorganización interna.

Por ello, cabe esperar que los cambios más significativos en la exposición territorial a la IA se produzcan en un horizonte de medio y largo plazo, en función de cómo evolucionen:

- la presencia de empresas digitales en cada territorio,
- la intensidad de adopción tecnológica en comercio, servicios y administración,
- los procesos de terciarización en curso en provincias actualmente industriales o agrarias,
- y las estrategias regionales de innovación, digitalización y atracción de talento.

En conjunto, la estabilidad entre 2021 y 2022 refuerza la idea de que la exposición a la IA en España es un fenómeno estructural, determinado por especializaciones productivas persistentes y distribuciones territoriales consolidadas. Por ello, los resultados deben interpretarse no como una radiografía coyuntural, sino como un mapa estructural de vulnerabilidad y oportunidad tecnológica en el mercado laboral español.

# 5. Conclusiones y trabajos futuros

El análisis desarrollado en este informe ofrece una primera aproximación sistemática al impacto potencial de la inteligencia artificial sobre el mercado laboral español, integrando dimensiones territoriales y de género que suelen quedar relegadas en la literatura. El ejercicio permite visualizar con claridad cómo la estructura productiva del país condiciona la exposición de cada territorio, y revela que la IA no afectará por igual a todas las provincias, ni de la misma manera a mujeres y hombres.

## 5.1. Conclusiones

Una de las principales conclusiones es que la exposición estimada responde fundamentalmente a patrones estructurales y no coyunturales. La relativa estabilidad del IA-share entre 2021 y 2022 confirma que la vulnerabilidad o capacidad de aprovechamiento de la IA depende ante todo del peso de los sectores intensivos en tareas cognitivas, administrativas o de interacción lingüística. Este hallazgo sitúa la cuestión en un marco de medio y largo plazo, más ligado a las estrategias de especialización productiva y a los procesos de transformación económica que a fluctuaciones anuales del empleo.

El análisis también evidencia que las áreas metropolitanas y las economías insulares presentan niveles especialmente elevados de exposición, vinculados a su concentración de actividades avanzadas y servicios orientados a la información. Por el contrario, territorios con mayor peso de agricultura, manufacturas tradicionales o construcción registran valores más moderados, reflejando perfiles productivos menos alineados con las capacidades de la IA generativa.

La dimensión de género introduce un matiz crucial: el empleo femenino se sitúa sistemáticamente en niveles más altos de exposición, en parte por su presencia en sectores como educación, sanidad, comercio y servicios administrativos. Esta diferencia no es marginal ni episódica: aparece en todas las provincias, con una consistencia llamativa. Aunque la IA puede mejorar la productividad y aliviar ciertas cargas de trabajo, también puede transformar tareas que hoy sostienen la participación femenina en el empleo. Reconocer esta asimetría será clave para diseñar políticas que prevengan nuevas brechas tecnológicas.

En conjunto, los resultados apuntan a un paisaje desigual, pero inteligible. La IA no se distribuye de manera caprichosa; sigue las líneas que ya trazan la especialización territorial, la división del trabajo por género y la configuración económica de cada provincia. Entender ese mapa es un paso previo indispensable para cualquier estrategia de preparación, adaptación o regulación.

## 5.1. Trabajos futuros

El carácter preliminar del estudio abre un conjunto amplio de líneas de trabajo que permitirán refinar, ampliar y profundizar los resultados obtenidos. Una primera prioridad consiste en avanzar desde el nivel sectorial al ocupacional. La disponibilidad creciente de clasificaciones ENOE, ESCO o microdatos de afiliación podría permitir una correspondencia más precisa entre sectores y tareas, reduciendo la incertidumbre asociada al uso de promedios sectoriales.

Otra línea de mejora pasa por incorporar dimensiones dinámicas. La metodología actual captura la estructura productiva, pero no la velocidad a la que empresas y administraciones están adoptando la IA. Integrar indicadores de automatización efectiva, inversión tecnológica, penetración de herramientas generativas o intensidad de digitalización permitiría estimar escenarios de transición más realistas.

También sería conveniente explorar efectos distributivos internos dentro de cada territorio: brechas por nivel educativo, edad, tamaño de empresa o tipo de contrato, que podrían interactuar con la exposición a la IA de formas no triviales. Del mismo modo, la construcción de índices complementarios —por ejemplo, de capacidad de adaptación o resiliencia territorial— permitiría matizar la lectura de la exposición bruta.

Desde la perspectiva territorial, futuras investigaciones podrían analizar cómo evolucionan los perfiles provinciales a la luz de las estrategias regulatorias europeas, las infraestructuras digitales o la implantación de hubs tecnológicos regionales. Asimismo, el uso de datos longitudinales permitiría observar si los territorios con mayor exposición son también los que incorporan más rápidamente prácticas de trabajo basadas en IA o si, por el contrario, se generan desigualdades adicionales.

Por último, la dimensión de género merece un tratamiento más profundo. Será crucial evaluar cómo interactúan las decisiones de adopción tecnológica con roles de cuidado, disponibilidad horaria, patrones de contratación y segregación ocupacional. Un enfoque interseccional —que incorpore edad, nivel formativo o situación migratoria— enriquecería aún más el análisis.

En conjunto, el informe sienta las bases para un programa de investigación más amplio que combine análisis estructural, evidencia empírica y reflexión estratégica. La IA no transformará el mercado laboral de manera uniforme; comprender quién está más expuesto y por qué es el primer paso para garantizar que la transición tecnológica sea inclusiva, territorialmente equilibrada y socialmente justa.

# Bibliografía

**[1]** D. Autor, F. Levy, and R. Murnane, “The Skill Content of Recent Technological Change,” *Quarterly Journal of Economics*, vol. 118, no. 4, pp. 1279–1333, 2003.

**[2]** E. Brynjolfsson and T. Mitchell, “What Can Machine Learning Do? Workforce Implications,” *Science*, vol. 358, pp. 1530–1534, 2017.

**[3]** E. Felten, M. Raj, and R. Seamans, “How Will Language Modelers Affect Occupations?,” *Brookings Papers on Economic Activity*, 2023.

**[4]** T. Tomlinson, J. Fernandez, S. Agarwal, et al., “Working with AI: Measuring the Applicability of Generative AI to Occupations,” 2025.

**[5]** T. Eloundou, S. Manning, P. Mishkin, and A. Rock, “GPTs are GPTs: An Early Look at the Labor Market Impact Potential of Large Language Models,” arXiv:2303.10130, 2023.

**[6]** L. Nurski, “The Exposure of EU Occupations to Generative AI,” *Bruegel Working Paper*, 2024.

**[7]** E. Brynjolfsson, Y. Li, and L. Raymond, “Generative AI at Work,” *NBER Working Paper* 31161, 2023.

**[8]** Banco de España, “El impacto de la inteligencia artificial en las tareas laborales,” *Informe Analítico*, 2024.

**[9]** European Central Bank, “Labour Market Effects of AI Adoption in Europe,” *ECB Working Paper*, 2024.

**[10]** D. Guarascio, L. Marcolin, and M. Squicciarini, “AI Exposure and Regional Labour Markets in Europe,” *Regional Studies*, 2023.

**[11]** OECD, “Local Employment Implications of AI Adoption,” *OECD Local Economic and Employment Development (LEED)*, 2023.

**[12]** OECD, “AI Ecosystems in Spain,” *OECD Digital Economy Papers*, 2024.

**[13]** OECD, “Algorithm and Eve: Gender and AI in Labour Markets,” 2023.

**[14]** European Central Bank, “AI Adoption and Female Employment Participation in Europe,” *ECB Working Paper*, 2024.

**[15]** International Labour Organization (ILO), “Generative AI and its Implications for Women’s Work,” *ILO Report*, 2023.

**[16]** Funcas, “Exposición tecnológica de las titulaciones universitarias en España,” *Documento de Trabajo*, 2024.

# ANEXO

## Anexo I: Mapas de incidencia IA

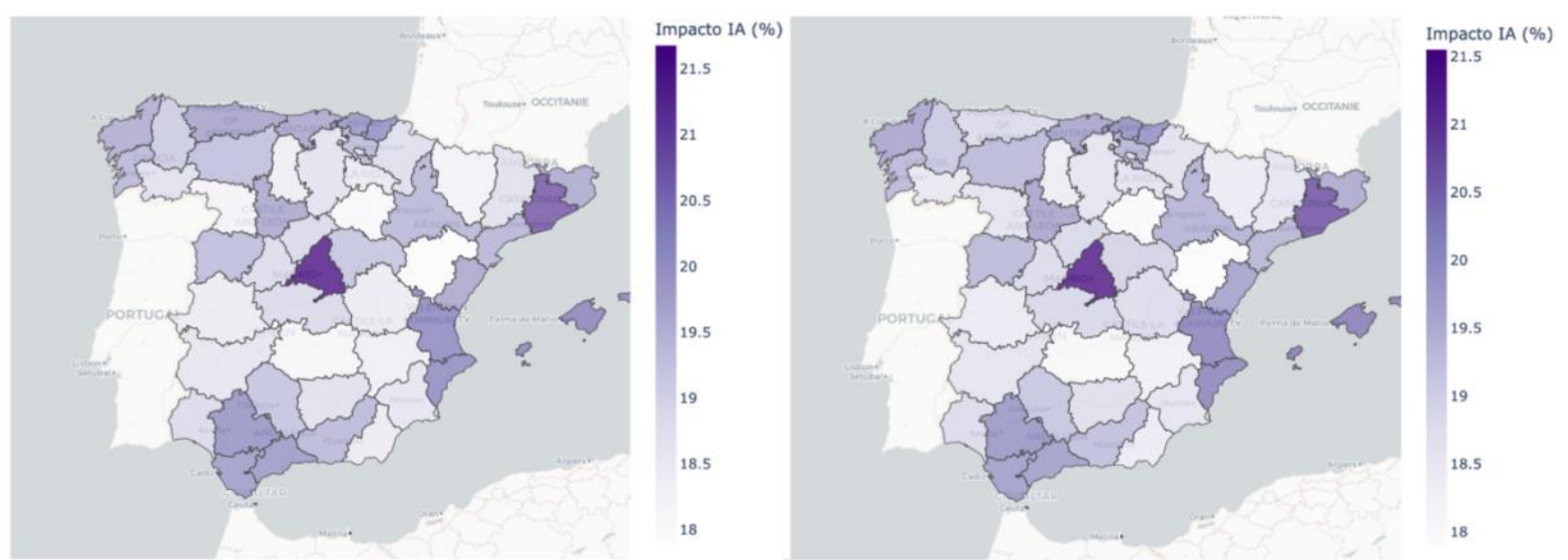

Incidencia de la IA en el trabajo en el 2022 (izquierda) y en el 2021 (derecha).

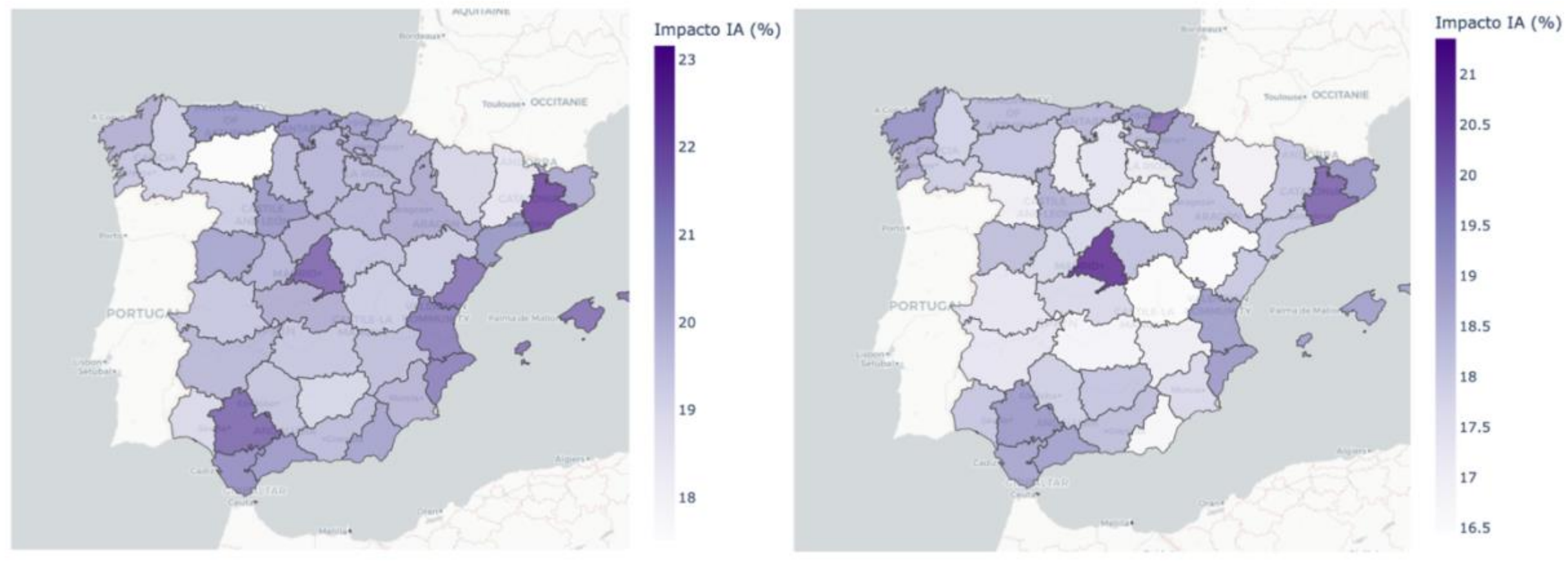

Incidencia de la IA en el trabajo en el 2022 en las mujeres (izquierda) y en los hombres (derecha).

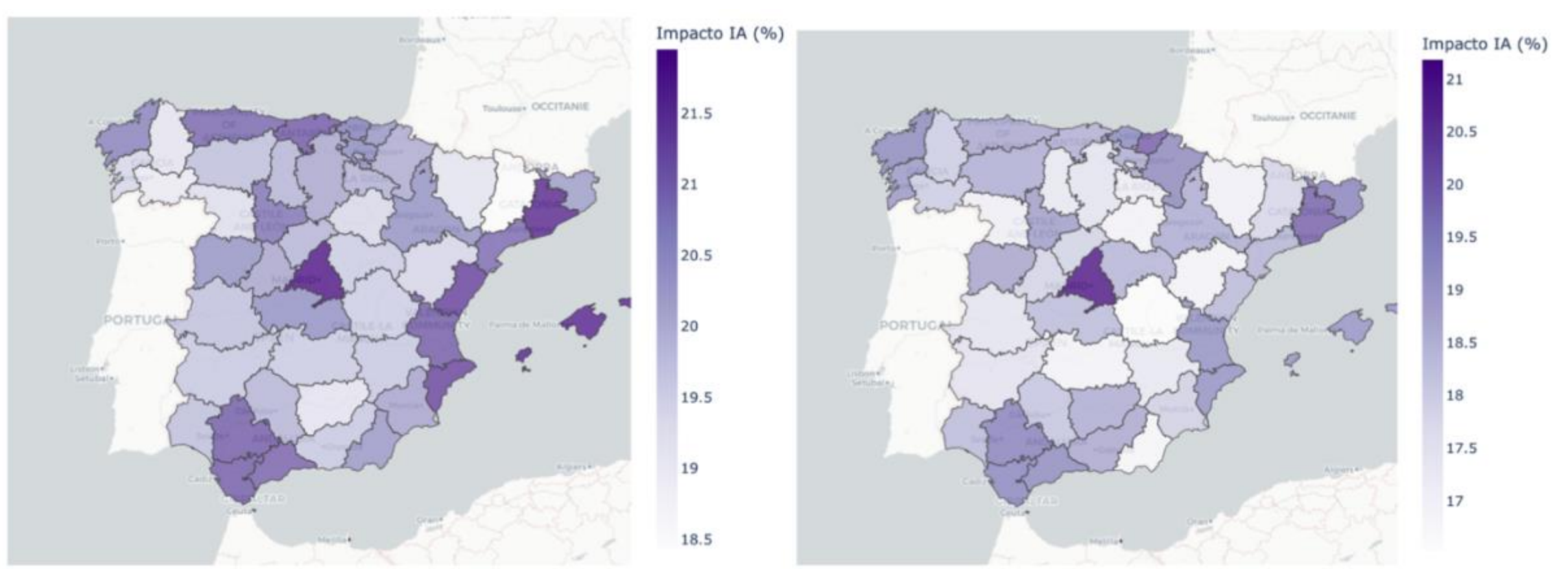

Incidencia de la IA en el trabajo en el 2021 en las mujeres (izquierda) y en los hombres (derecha).

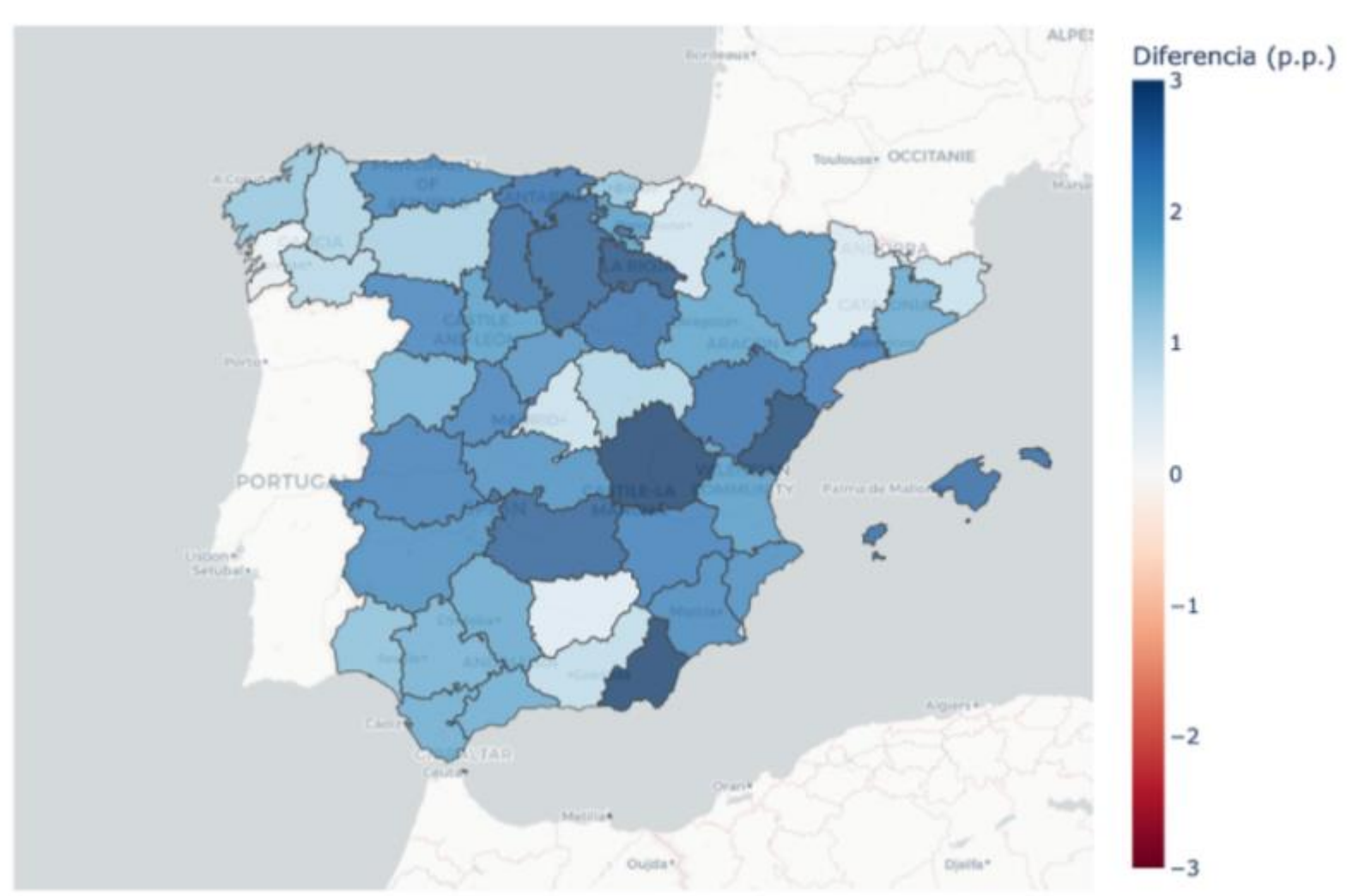

Diferencia de la incidencia de la IA en el trabajo entre mujeres (azul) y hombres (rojo).

## Anexo II: Matriz de incidencia IA

# Sectores primarios y extractivos

| CNAE | Factor IA | Justificación |
|---|---|---|
| 01 Agricultura | 0,060 | Tareas físicas y rutinarias; grupo *Farming*. |
| 02 Silvicultura | 0,060 | Ídem 01. |
| 03 Pesca y acuicultura | 0,060 | Actividades físicas y manuales. |
| 05 Extracción carbón | 0,080 | Similar a *Construction & Extraction*. |
| 06 Extracción petróleo y gas | 0,080 | Tareas operativas y mecánicas. |
| 07 Minerales metálicos | 0,080 | Exposición limitada a IA generativa. |
| 08 Otras industrias extractivas | 0,080 | Perfil ocupacional operativo. |
| 09 Apoyo a extractivas | 0,080 | Actividades auxiliares manuales. |

---

# Manufactura y producción tradicional

| CNAE | Factor IA | Justificación |
|---|---|---|
| 10 Alimentación | 0,145 | Mezcla de *Production* (0,11) + *Food Service* (0,18). |
| 11 Bebidas | 0,145 | Ídem 10. |
| 12 Tabaco | 0,110 | Manufactura clásica. |
| 13 Textil | 0,110 | *Production* manual. |
| 14 Confección | 0,110 | Patronaje y ensamblaje manual. |
| 15 Cuero y calzado | 0,110 | *Production*. |
| 16 Madera y corcho | 0,110 | Manufactura rutinaria. |
| 17 Papel | 0,110 | Producción industrial. |
| 18 Artes gráficas | 0,110 | Tareas técnicas pero manuales. |
| 19 Coquerías y refino | 0,165 | *Production* + *Engineering*. |
| 20 Química | 0,165 | Actividad técnica y de laboratorio. |
| 21 Farmacéutica | 0,165 | I+D + manufactura avanzada. |
| 22 Caucho y plástico | 0,165 | Procesos técnico-industriales. |
| 23 Minerales no metálicos | 0,165 | Industria intensiva en ingeniería. |
| 24 Metalurgia | 0,165 | Procesos técnicos de precisión. |
| 25 Productos metálicos | 0,110 | Manufactura estándar. |
| 26 Informáticos / electrónicos | 0,260 | *Computer & Math* (0,30) + *Engineering* (0,22). |
| 27 Eléctrico | 0,165 | Técnico-industrial. |
| 28 Maquinaria | 0,165 | Ingeniería + producción avanzada. |
| 29 Automoción | 0,165 | Ingeniería aplicada. |
| 30 Transporte (fabricación) | 0,165 | Ingeniería + producción compleja. |

| CNAE | Factor IA | Justificación |
|---|---|---|
| 31 Muebles | 0,110 | Manufactura tradicional. |
| 32 Otras manufactureras | 0,110 | *Production*. |
| 33 Reparación e instalación | 0,110 | *Installation, Maintenance & Repair*. |

---

## Energía, medio ambiente y construcción

| CNAE | Factor IA | Justificación |
|---|---|---|
| 35 Energía | 0,110 | Tareas técnicas + operativas. |
| 36 Agua | 0,110 | Supervisión operativa. |
| 37 Aguas residuales | 0,110 | Mantenimiento / operativa. |
| 38 Residuos | 0,110 | Actividades manuales. |
| 39 Descontaminación | 0,110 | Actividades técnicas repetitivas. |
| 41 Edificación | 0,095 | Mix *Construction* (0,08) + *Maintenance* (0,11). |
| 42 Ingeniería civil | 0,170 | Incluye perfiles de ingeniería. |
| 43 Construcción especializada | 0,095 | Predominio de tareas físicas. |

---

## Comercio, logística y transporte

| CNAE | Factor IA | Justificación |
|---|---|---|
| 45 Vehículos | 0,305 | *Sales* + *Office/Admin*. |
| 46 Mayorista | 0,305 | Actividad comercial intensiva en gestión. |
| 47 Minorista | 0,305 | Interacción lingüística constante. |
| 49 Transporte terrestre | 0,110 | Tareas físicas + logística básica. |
| 50 Transporte marítimo | 0,110 | Predominio operativo. |
| 51 Transporte aéreo | 0,110 | Procedimientos estandarizados. |
| 52 Almacenamiento | 0,200 | Logística + administración. |
| 53 Correos | 0,200 | Clasificación + tareas administrativas. |

---

## Turismo, hostelería y servicios personales

| CNAE | Factor IA | Justificación |
|---|---|---|
| 55 Alojamiento | 0,250 | *Sales* + *Food Service* + administración. |
| 56 Comidas y bebidas | 0,250 | Interacción + logística + tareas repetitivas. |

---

## Sectores culturales, informacionales y comunicativos

| CNAE | Factor IA | Justificación |
|---|---|---|
| 58 Edición | 0,250 | Tareas de redacción y procesamiento lingüístico. |
| 59 Audiovisual | 0,250 | Edición, guion, producción. |

| CNAE | Factor IA | Justificación |
| --- | --- | --- |
| 60 Radio y TV | 0,250 | Creación de contenidos + documentación. |
| 61 Telecomunicaciones | 0,300 | Ingeniería + datos + información. |
| 62 Informática y consultoría | 0,300 | *Computer & Math* pleno. |
| 63 Servicios de información | 0,300 | Tareas de análisis y síntesis informativa. |

---

## Actividades financieras e inmobiliarias

| CNAE | Factor IA | Justificación |
| --- | --- | --- |
| 64 Servicios financieros | 0,280 | *Business & Financial* + *Admin*. |
| 65 Seguros | 0,280 | Procesos normativos y analíticos. |
| 66 Auxiliares financieros | 0,290 | Tareas administrativas complejas. |
| 68 Inmobiliarias | 0,280 | *Sales* + *Financial operations*. |

---

## Profesionales, científicas, técnicas y auxiliares

| CNAE | Factor IA | Justificación |
| --- | --- | --- |
| 69 Jurídicas y contabilidad | 0,240 | Documentación y análisis. |
| 70 Consultoría gestión | 0,260 | Asesoría, análisis, informes. |
| 71 Ingeniería y arquitectura | 0,260 | Modelado, planificación, cálculos. |
| 72 I+D | 0,250 | Análisis y síntesis científica. |
| 73 Publicidad | 0,250 | Creatividad + gestión informativa. |
| 74 Otras profesionales | 0,240 | Alta carga cognitiva. |
| 75 Veterinarias | 0,150 | Procesos clínicos y técnicos. |
| 77 Alquiler | 0,260 | Gestión documental. |
| 78 Empleo | 0,260 | Intermediación laboral + administración. |
| 79 Agencias de viaje | 0,260 | Planificación + atención cliente. |
| 80 Seguridad privada | 0,200 | Monitorización + procedimientos. |
| 81 Jardinería y edificios | 0,110 | Tareas operativas. |
| 82 Administrativas y apoyo | 0,290 | *Office & Admin* puro. |

---

## Administración pública, educación y salud

| CNAE | Factor IA | Justificación |
| --- | --- | --- |
| 84 Administración pública | 0,195 | *Management* + *Social service*. |
| 85 Educación | 0,230 | Actividades docentes y administrativas. |
| 86 Sanidad | 0,120 | *Healthcare* técnico. |
| 87 Residencias | 0,150 | Cuidado + administración. |
| 88 Servicios sociales | 0,150 | Tareas asistenciales. |

---

## Cultura, ocio y servicios comunitarios

| CNAE | Factor IA | Justificación |
| --- | --- | --- |
| 90 Artes y espectáculos | 0,230 | Creación + gestión informativa. |
| 91 Bibliotecas y museos | 0,250 | Curaduría + documentación. |
| 92 Juegos de azar | 0,230 | Gestión + operaciones. |
| 93 Deportes | 0,220 | Administración + servicios. |
| 94 Asociaciones | 0,250 | Gestión documental. |
| 95 Reparación | 0,110 | Reparación técnica manual. |
| 96 Servicios personales | 0,200 | Atención + servicios. |
| 97 Hogares empleadores | 0,200 | Tareas no cognitivas. |
| 98 Hogares productores | 0,200 | Actividades manuales. |
| 99 Organismos extraterritoriales | 0,140 | Gestión directiva. |

## Anexo III: Incidencia IA sobre el total 2022

| Provincia | Empleo total | Empleo IA | IA-share |
|---|---|---|---|
| 01 Araba/Álava | 142.789 | 27.818 | **0,1948** |
| 02 Albacete | 160.804 | 29.534 | **0,1837** |
| 03 Alicante | 698.158 | 141.648 | **0,2029** |
| 04 Almería | 298.738 | 55.237 | **0,1849** |
| 05 Ávila | 63.277 | 11.970 | **0,1892** |
| 06 Badajoz | 257.720 | 48.052 | **0,1865** |
| 07 Illes Balears | 467.233 | 95.316 | **0,204** |
| 08 Barcelona | 2.475.590 | 517.127 | **0,2089** |
| 09 Burgos | 150.488 | 28.180 | **0,1873** |
| 10 Cáceres | 148.769 | 27.471 | **0,1847** |
| 11 Cádiz | 410.949 | 82.252 | **0,2002** |
| 12 Castellón | 247.936 | 49.170 | **0,1983** |
| 13 Ciudad Real | 194.797 | 35.291 | **0,1812** |
| 14 Córdoba | 309.610 | 59.911 | **0,1935** |
| 15 A Coruña | 385.483 | 76.080 | **0,1973** |
| 16 Cuenca | 76.653 | 14.159 | **0,1847** |
| 17 Girona | 379.389 | 75.005 | **0,1977** |
| 18 Granada | 357.867 | 69.896 | **0,1953** |
| 19 Guadalajara | 130.380 | 25.201 | **0,1932** |
| 20 Gipuzkoa | 287.294 | 58.018 | **0,2019** |
| 21 Huelva | 210.901 | 39.931 | **0,1893** |
| 22 Huesca | 112.931 | 20.640 | **0,1829** |
| 23 Jaén | 211.782 | 39.780 | **0,1879** |
| 24 León | 188.858 | 36.682 | **0,1942** |
| 25 Lleida | 195.192 | 36.661 | **0,1878** |
| 26 La Rioja | 134.599 | 24.876 | **0,1847** |
| 27 Lugo | 112.944 | 21.706 | **0,1922** |
| 28 Madrid | 3.014.953 | 653.696 | **0,2168** |
| 29 Málaga | 665.807 | 133.583 | **0,2006** |
| 30 Murcia | 586.741 | 109.490 | **0,1867** |
| 31 Navarra | 276.597 | 52.114 | **0,1885** |
| 32 Ourense | 105.511 | 19.840 | **0,1881** |
| 33 Asturias | 370.728 | 73.723 | **0,199** |
| 34 Palencia | 63.481 | 11.703 | **0,1843** |
| 35 Las Palmas | 386.306 | 81.422 | **0,2108** |
| 36 Pontevedra | 421.238 | 82.290 | **0,1954** |
| 37 Salamanca | 129.903 | 25.293 | **0,1947** |
| 38 Santa Cruz de Tenerife | 416.670 | 87.698 | **0,2105** |
| 39 Cantabria | 235.535 | 46.665 | **0,1981** |
| 40 Segovia | 68.326 | 12.955 | **0,1896** |
| 41 Sevilla | 733.484 | 147.797 | **0,2015** |
| 42 Soria | 39.124 | 7.052 | **0,1803** |
| 43 Tarragona | 332.867 | 65.209 | **0,1959** |
| 44 Teruel | 57.348 | 10.267 | **0,179** |
| 45 Toledo | 296.489 | 56.217 | **0,1896** |
| 46 Valencia | 1.083.550 | 220.091 | **0,2031** |
| 47 Valladolid | 220.675 | 43.694 | **0,198** |
| 48 Bizkaia | 474.815 | 95.544 | **0,2012** |
| 49 Zamora | 65.791 | 11.981 | **0,1821** |
| 50 Zaragoza | 421.764 | 82.476 | **0,1956** |
| 51 Ceuta | 28.369 | 5.776 | **0,2036** |
| 52 Melilla | 29.888 | 6.075 | **0,2033** |

## Anexo IV: Incidencia IA sobre el total 2021

| Provincia | Empleo total | Empleo IA (aprox) | % IA-share |
|---|---|---|---|
| 01 Araba/Álava | 137.916 | 26.977 | **0,1956** |
| 02 Albacete | 153.694 | 28.182 | **0,1834** |
| 03 Alicante | 658.787 | 133.720 | **0,203** |
| 04 Almería | 288.223 | 53.278 | **0,1849** |
| 05 Ávila | 60.731 | 11.469 | **0,1889** |
| 06 Badajoz | 248.124 | 46.291 | **0,1866** |
| 07 Illes Balears | 443.628 | 90.552 | **0,2041** |
| 08 Barcelona | 2.383.924 | 498.004 | **0,2089** |
| 09 Burgos | 146.255 | 27.333 | **0,1869** |
| 10 Cáceres | 143.786 | 26.588 | **0,1849** |
| 11 Cádiz | 390.799 | 77.929 | **0,1994** |
| 12 Castellón | 239.073 | 47.509 | **0,1987** |
| 13 Ciudad Real | 187.679 | 33.906 | **0,1807** |
| 14 Córdoba | 299.707 | 57.775 | **0,1927** |
| 15 A Coruña | 368.981 | 73.046 | **0,1981** |
| 16 Cuenca | 76.196 | 14.382 | **0,1887** |
| 17 Girona | 363.553 | 71.796 | **0,1974** |
| 18 Granada | 347.482 | 67.415 | **0,1941** |
| 19 Guadalajara | 124.551 | 23.716 | **0,1905** |
| 20 Gipuzkoa | 281.226 | 56.585 | **0,2012** |
| 21 Huelva | 200.942 | 37.948 | **0,1888** |
| 22 Huesca | 107.965 | 19.983 | **0,1851** |
| 23 Jaén | 204.917 | 38.550 | **0,1882** |
| 24 León | 178.230 | 34.737 | **0,1948** |
| 25 Lleida | 184.630 | 34.395 | **0,1862** |
| 26 La Rioja | 125.568 | 23.474 | **0,187** |
| 27 Lugo | 107.270 | 20.632 | **0,1924** |
| 28 Madrid | 2.906.065 | 626.297 | **0,2157** |
| 29 Málaga | 646.256 | 128.796 | **0,1993** |
| 30 Murcia | 556.100 | 103.870 | **0,1869** |
| 31 Navarra | 259.823 | 48.664 | **0,1873** |
| 32 Ourense | 100.386 | 18.620 | **0,1855** |
| 33 Asturias | 355.985 | 67.295 | **0,189** |
| 34 Palencia | 60.928 | 11.233 | **0,1844** |
| 35 Las Palmas | 367.617 | 77.333 | **0,2104** |
| 36 Pontevedra | 400.336 | 77.985 | **0,1947** |
| 37 Salamanca | 126.073 | 24.579 | **0,195** |
| 38 Santa Cruz de Tenerife | 391.222 | 82.396 | **0,2106** |

| 39 Cantabria | 227.140 | 45.022 | **0,1982** |
|---|---|---|---|
| 40 Segovia | 65.721 | 12.441 | **0,1893** |
| 41 Sevilla | 704.316 | 141.197 | **0,2005** |
| 42 Soria | 38.521 | 6.918 | **0,1796** |
| 43 Tarragona | 316.839 | 61.925 | **0,1954** |
| 44 Teruel | 55.229 | 9.886 | **0,179** |
| 45 Toledo | 282.255 | 53.484 | **0,1895** |
| 46 Valencia | 1.042.481 | 211.636 | **0,203** |
| 47 Valladolid | 214.866 | 42.563 | **0,1981** |
| 48 Bizkaia | 465.681 | 93.872 | **0,2016** |
| 49 Zamora | 64.152 | 11.652 | **0,1816** |
| 50 Zaragoza | 408.177 | 79.926 | **0,1958** |
| 51 Ceuta | 28.050 | 5.742 | **0,2047** |
| 52 Melilla | 29.783 | 6.058 | **0,2034** |

## Anexo V: Incidencia IA sobre el empleo en mujeres 2022

| Provincia | Empleo mujeres | Empleo IA-aplic. | IA-share |
|---|---|---|---|
| **01 Araba/Álava** | 68.127 | 13.945 | **0,2047** |
| **02 Albacete** | 70.611 | 14.067 | **0,1992** |
| **03 Alicante** | 319.236 | 68.348 | **0,2141** |
| **04 Almería** | 133.171 | 27.485 | **0,2064** |
| **05 Ávila** | 28.231 | 5.687 | **0,2014** |
| **06 Badajoz** | 112.997 | 22.692 | **0,2008** |
| **07 Illes Balears** | 215.956 | 46.946 | **0,2174** |
| **08 Barcelona** | 1.192.178 | 256.308 | **0,215** |
| **09 Burgos** | 69.033 | 13.898 | **0,2013** |
| **10 Cáceres** | 67.238 | 13.246 | **0,197** |
| **11 Cádiz** | 178.617 | 37.821 | **0,2117** |
| **12 Castellón** | 114.349 | 24.709 | **0,2161** |
| **13 Ciudad Real** | 84.883 | 16.754 | **0,1974** |
| **14 Córdoba** | 148.321 | 29.311 | **0,1976** |
| **15 A Coruña** | 189.360 | 38.462 | **0,2031** |
| **16 Cuenca** | 35.670 | 6.977 | **0,1957** |
| **17 Girona** | 183.522 | 37.527 | **0,2045** |
| **18 Granada** | 175.346 | 35.039 | **0,1998** |
| **19 Guadalajara** | 61.319 | 12.113 | **0,1975** |
| **20 Gipuzkoa** | 160.032 | 33.229 | **0,2077** |
| **21 Huelva** | 90.402 | 17.803 | **0,1969** |
| **22 Huesca** | 49.695 | 9.582 | **0,1927** |
| **23 Jaén** | 98.191 | 18.884 | **0,1923** |
| **24 León** | 90.996 | 17.835 | **0,196** |
| **25 Lleida** | 87.076 | 16.341 | **0,1877** |
| **26 La Rioja** | 65.324 | 13.069 | **0,2001** |
| **27 Lugo** | 45.766 | 8.894 | **0,1943** |
| **28 Madrid** | 1.474.951 | 323.634 | **0,2194** |
| **29 Málaga** | 324.669 | 67.968 | **0,2094** |
| **30 Murcia** | 270.205 | 54.581 | **0,202** |
| **31 Navarra** | 135.176 | 27.149 | **0,2008** |
| **32 Ourense** | 43.011 | 8.314 | **0,1934** |
| **33 Asturias** | 186.251 | 39.008 | **0,2094** |
| **34 Palencia** | 29.995 | 5.986 | **0,1996** |
| **35 Las Palmas** | 194.685 | 41.500 | **0,2132** |
| **36 Pontevedra** | 203.709 | 40.208 | **0,1974** |
| **37 Salamanca** | 59.790 | 12.291 | **0,2056** |
| **38 Santa Cruz de Tenerife** | 200.032 | 43.923 | **0,2196** |
| **39 Cantabria** | 110.823 | 23.372 | **0,2109** |
| **40 Segovia** | 30.663 | 6.228 | **0,2031** |
| **41 Sevilla** | 331.468 | 69.986 | **0,2111** |
| **42 Soria** | 17.857 | 3.472 | **0,1944** |
| **43 Tarragona** | 153.435 | 32.218 | **0,21** |
| **44 Teruel** | 25.785 | 5.045 | **0,1957** |
| **45 Toledo** | 128.547 | 26.185 | **0,2037** |
| **46 Valencia** | 506.749 | 108.777 | **0,2147** |
| **47 Valladolid** | 103.781 | 21.715 | **0,2092** |
| **48 Bizkaia** | 230.311 | 48.554 | **0,2108** |
| **49 Zamora** | 29.721 | 5.814 | **0,1956** |
| **50 Zaragoza** | 198.316 | 41.065 | **0,2071** |
| **51 Ceuta** | 11.358 | 2.382 | **0,2098** |
| **52 Melilla** | 12.316 | 2.524 | **0,205** |

## Anexo VI: Incidencia IA sobre el empleo en hombres 2022

| Provincia | Empleo hombres | Empleo IA-aplicado | IA-share |
|---|---|---|---|
| **01 Araba/Álava** | 74.662 | 13.873 | **0,1858** |
| **02 Albacete** | 90.193 | 15.468 | **0,1715** |
| **03 Alicante** | 378.922 | 73.300 | **0,1934** |
| **04 Almería** | 165.567 | 27.753 | **0,1676** |
| **05 Ávila** | 35.046 | 6.283 | **0,1793** |
| **06 Badajoz** | 144.723 | 25.360 | **0,1752** |
| **07 Illes Balears** | 251.277 | 48.370 | **0,1925** |
| **08 Barcelona** | 1.283.412 | 260.819 | **0,2032** |
| **09 Burgos** | 81.455 | 14.281 | **0,1753** |
| **10 Cáceres** | 81.531 | 14.225 | **0,1745** |
| **11 Cádiz** | 232.332 | 44.430 | **0,1912** |
| **12 Castellón** | 133.587 | 24.461 | **0,1831** |
| **13 Ciudad Real** | 109.914 | 18.537 | **0,1687** |
| **14 Córdoba** | 161.289 | 29.258 | **0,1814** |
| **15 A Coruña** | 196.123 | 38.327 | **0,1955** |
| **16 Cuenca** | 41.006 | 6.751 | **0,1646** |
| **17 Girona** | 195.867 | 38.185 | **0,195** |
| **18 Granada** | 182.521 | 33.857 | **0,1855** |
| **19 Guadalajara** | 69.061 | 12.734 | **0,1844** |
| **20 Gipuzkoa** | 157.994 | 31.807 | **0,2012** |
| **21 Huelva** | 120.499 | 22.127 | **0,1837** |
| **22 Huesca** | 63.236 | 10.778 | **0,1704** |
| **23 Jaén** | 113.591 | 20.896 | **0,1839** |
| **24 León** | 97.862 | 18.006 | **0,184** |
| **25 Lleida** | 108.116 | 19.833 | **0,1835** |
| **26 La Rioja** | 74.299 | 12.801 | **0,1723** |
| **27 Lugo** | 67.178 | 12.139 | **0,1807** |
| **28 Madrid** | 1.541.257 | 326.429 | **0,2118** |
| **29 Málaga** | 341.138 | 65.615 | **0,1923** |
| **30 Murcia** | 316.297 | 56.612 | **0,179** |
| **31 Navarra** | 141.240 | 26.987 | **0,1911** |
| **32 Ourense** | 62.500 | 11.387 | **0,1822** |
| **33 Asturias** | 209.466 | 39.008 | **0,1864** |
| **34 Palencia** | 33.486 | 5.770 | **0,1723** |
| **35 Las Palmas** | 191.621 | 40.922 | **0,2135** |
| **36 Pontevedra** | 217.529 | 41.082 | **0,1888** |
| **37 Salamanca** | 70.113 | 13.002 | **0,1854** |
| **38 Santa Cruz de Tenerife** | 216.638 | 43.775 | **0,2021** |
| **39 Cantabria** | 124.712 | 23.293 | **0,1868** |
| **40 Segovia** | 37.663 | 6.727 | **0,1786** |
| **41 Sevilla** | 402.016 | 77.812 | **0,1936** |
| **42 Soria** | 21.267 | 3.580 | **0,1683** |
| **43 Tarragona** | 179.432 | 32.991 | **0,1839** |
| **44 Teruel** | 31.563 | 5.222 | **0,1654** |
| **45 Toledo** | 167.942 | 30.031 | **0,1788** |
| **46 Valencia** | 576.801 | 111.313 | **0,193** |
| **47 Valladolid** | 116.894 | 21.979 | **0,188** |
| **48 Bizkaia** | 244.504 | 46.990 | **0,1922** |
| **49 Zamora** | 36.070 | 6.167 | **0,171** |
| **50 Zaragoza** | 223.448 | 41.412 | **0,1853** |
| **51 Ceuta** | 17.011 | 3.394 | **0,1995** |
| **52 Melilla** | 17.572 | 3.551 | **0,2021** |

## Anexo VII: Incidencia IA sobre el empleo en mujeres 2021

| Provincia | Empleo mujeres | Empleo IA-aplicado | IA-share |
|---|---|---|---|
| **01 Araba/Álava** | 63.254 | 13.012 | **0,2057** |
| **02 Albacete** | 63.501 | 12.503 | **0,1967** |
| **03 Alicante** | 303.895 | 65.008 | **0,214** |
| **04 Almería** | 122.656 | 25.023 | **0,2041** |
| **05 Ávila** | 25.685 | 5.184 | **0,2018** |
| **06 Badajoz** | 108.981 | 21.441 | **0,1967** |
| **07 Illes Balears** | 204.288 | 44.516 | **0,2179** |
| **08 Barcelona** | 1.140.512 | 247.802 | **0,2173** |
| **09 Burgos** | 64.800 | 13.062 | **0,2015** |
| **10 Cáceres** | 62.255 | 12.322 | **0,1979** |
| **11 Cádiz** | 168.467 | 35.604 | **0,2113** |
| **12 Castellón** | 112.428 | 24.089 | **0,2142** |
| **13 Ciudad Real** | 77.765 | 15.264 | **0,1963** |
| **14 Córdoba** | 138.418 | 27.629 | **0,1996** |
| **15 A Coruña** | 177.858 | 36.864 | **0,2073** |
| **16 Cuenca** | 34.521 | 6.767 | **0,196** |
| **17 Girona** | 167.686 | 34.055 | **0,2031** |
| **18 Granada** | 165.349 | 32.552 | **0,1968** |
| **19 Guadalajara** | 55.563 | 10.899 | **0,1962** |
| **20 Gipuzkoa** | 153.232 | 31.293 | **0,2043** |
| **21 Huelva** | 87.319 | 17.308 | **0,1982** |
| **22 Huesca** | 44.729 | 8.575 | **0,1918** |
| **23 Jaén** | 91.326 | 17.529 | **0,1919** |
| **24 León** | 80.368 | 15.939 | **0,1983** |
| **25 Lleida** | 76.514 | 14.104 | **0,1843** |
| **26 La Rioja** | 62.551 | 12.514 | **0,2001** |
| **27 Lugo** | 44.095 | 8.444 | **0,1915** |
| **28 Madrid** | 1.415.473 | 310.717 | **0,2195** |
| **29 Málaga** | 305.118 | 64.291 | **0,2107** |
| **30 Murcia** | 256.360 | 51.790 | **0,202** |
| **31 Navarra** | 130.180 | 26.179 | **0,201** |
| **32 Ourense** | 40.924 | 7.784 | **0,1902** |
| **33 Asturias** | 180.324 | 37.879 | **0,21** |
| **34 Palencia** | 27.442 | 5.478 | **0,1996** |
| **35 Las Palmas** | 176.002 | 36.063 | **0,2049** |
| **36 Pontevedra** | 196.831 | 38.241 | **0,1943** |
| **37 Salamanca** | 56.417 | 11.550 | **0,2046** |
| **38 Santa Cruz de Tenerife** | 215.698 | 46.333 | **0,2147** |

| **39 Cantabria** | 106.560 | 22.527 | **0,2114** |
|---|---|---|---|
| **40 Segovia** | 29.240 | 5.833 | **0,1996** |
| **41 Sevilla** | 319.923 | 67.607 | **0,2114** |
| **42 Soria** | 17.254 | 3.350 | **0,1942** |
| **43 Tarragona** | 143.404 | 30.059 | **0,2096** |
| **44 Teruel** | 24.702 | 4.804 | **0,1945** |
| **45 Toledo** | 120.829 | 24.756 | **0,205** |
| **46 Valencia** | 483.403 | 102.165 | **0,2114** |
| **47 Valladolid** | 101.085 | 21.085 | **0,2086** |
| **48 Bizkaia** | 221.950 | 46.007 | **0,2072** |
| **49 Zamora** | 28.082 | 5.419 | **0,1929** |
| **50 Zaragoza** | 198.895 | 40.741 | **0,2047** |
| **51 Ceuta** | 11.260 | 2.381 | **0,2114** |
| **52 Melilla** | 12.248 | 2.507 | **0,2047** |

## Anexo VIII: Incidencia IA sobre el empleo en hombres 2021

| Provincia | Empleo hombres | Empleo IA-aplicado | IA-share |
|---|---|---|---|
| **01 Araba/Álava** | 74.662 | 13.965 | **0,187** |
| **02 Albacete** | 90.193 | 15.679 | **0,1738** |
| **03 Alicante** | 378.922 | 73.018 | **0,1927** |
| **04 Almería** | 165.567 | 27.942 | **0,1687** |
| **05 Ávila** | 35.046 | 6.285 | **0,1793** |
| **06 Badajoz** | 144.723 | 25.427 | **0,1757** |
| **07 Illes Balears** | 239.340 | 46.036 | **0,1924** |
| **08 Barcelona** | 1.243.412 | 249.202 | **0,2004** |
| **09 Burgos** | 81.455 | 14.271 | **0,1752** |
| **10 Cáceres** | 81.531 | 14.266 | **0,1751** |
| **11 Cádiz** | 222.332 | 43.346 | **0,1949** |
| **12 Castellón** | 126.645 | 23.426 | **0,185** |
| **13 Ciudad Real** | 109.914 | 18.648 | **0,1697** |
| **14 Córdoba** | 161.289 | 29.479 | **0,1829** |
| **15 A Coruña** | 191.123 | 37.193 | **0,1946** |
| **16 Cuenca** | 41.006 | 6.780 | **0,1653** |
| **17 Girona** | 195.867 | 38.260 | **0,1953** |
| **18 Granada** | 182.521 | 34.366 | **0,1883** |
| **19 Guadalajara** | 69.061 | 12.833 | **0,1858** |
| **20 Gipuzkoa** | 157.994 | 31.700 | **0,2008** |
| **21 Huelva** | 120.499 | 22.245 | **0,1846** |
| **22 Huesca** | 63.236 | 10.800 | **0,1708** |
| **23 Jaén** | 113.591 | 21.274 | **0,1872** |
| **24 León** | 97.862 | 18.314 | **0,1872** |
| **25 Lleida** | 108.116 | 19.325 | **0,1788** |
| **26 La Rioja** | 72.005 | 12.361 | **0,1717** |
| **27 Lugo** | 67.178 | 12.155 | **0,1809** |
| **28 Madrid** | 1.489.542 | 315.580 | **0,2119** |
| **29 Málaga** | 341.138 | 66.273 | **0,1943** |
| **30 Murcia** | 316.297 | 56.969 | **0,1801** |
| **31 Navarra** | 141.240 | 27.457 | **0,1943** |
| **32 Ourense** | 62.500 | 11.306 | **0,181** |
| **33 Asturias** | 209.466 | 39.416 | **0,1882** |
| **34 Palencia** | 33.486 | 5.820 | **0,1738** |
| **35 Las Palmas** | 191.621 | 40.598 | **0,212** |
| **36 Pontevedra** | 217.529 | 41.610 | **0,1913** |
| **37 Salamanca** | 70.113 | 13.244 | **0,1888** |
| **38 Santa Cruz de Tenerife** | 216.638 | 43.622 | **0,2013** |
| **39 Cantabria** | 120.580 | 22.495 | **0,1865** |
| **40 Segovia** | 37.663 | 6.759 | **0,1795** |
| **41 Sevilla** | 402.016 | 78.643 | **0,1957** |
| **42 Soria** | 21.267 | 3.603 | **0,1694** |
| **43 Tarragona** | 179.432 | 33.378 | **0,186** |
| **44 Teruel** | 31.563 | 5.353 | **0,1696** |
| **45 Toledo** | 167.942 | 30.986 | **0,1845** |
| **46 Valencia** | 576.801 | 111.079 | **0,1927** |
| **47 Valladolid** | 116.894 | 22.222 | **0,1901** |
| **48 Bizkaia** | 244.504 | 47.426 | **0,194** |
| **49 Zamora** | 36.070 | 6.168 | **0,171** |
| **50 Zaragoza** | 223.448 | 41.831 | **0,1872** |
| **51 Ceuta** | 16.790 | 3.361 | **0,2002** |
| **52 Melilla** | 17.535 | 3.551 | **0,2023** |

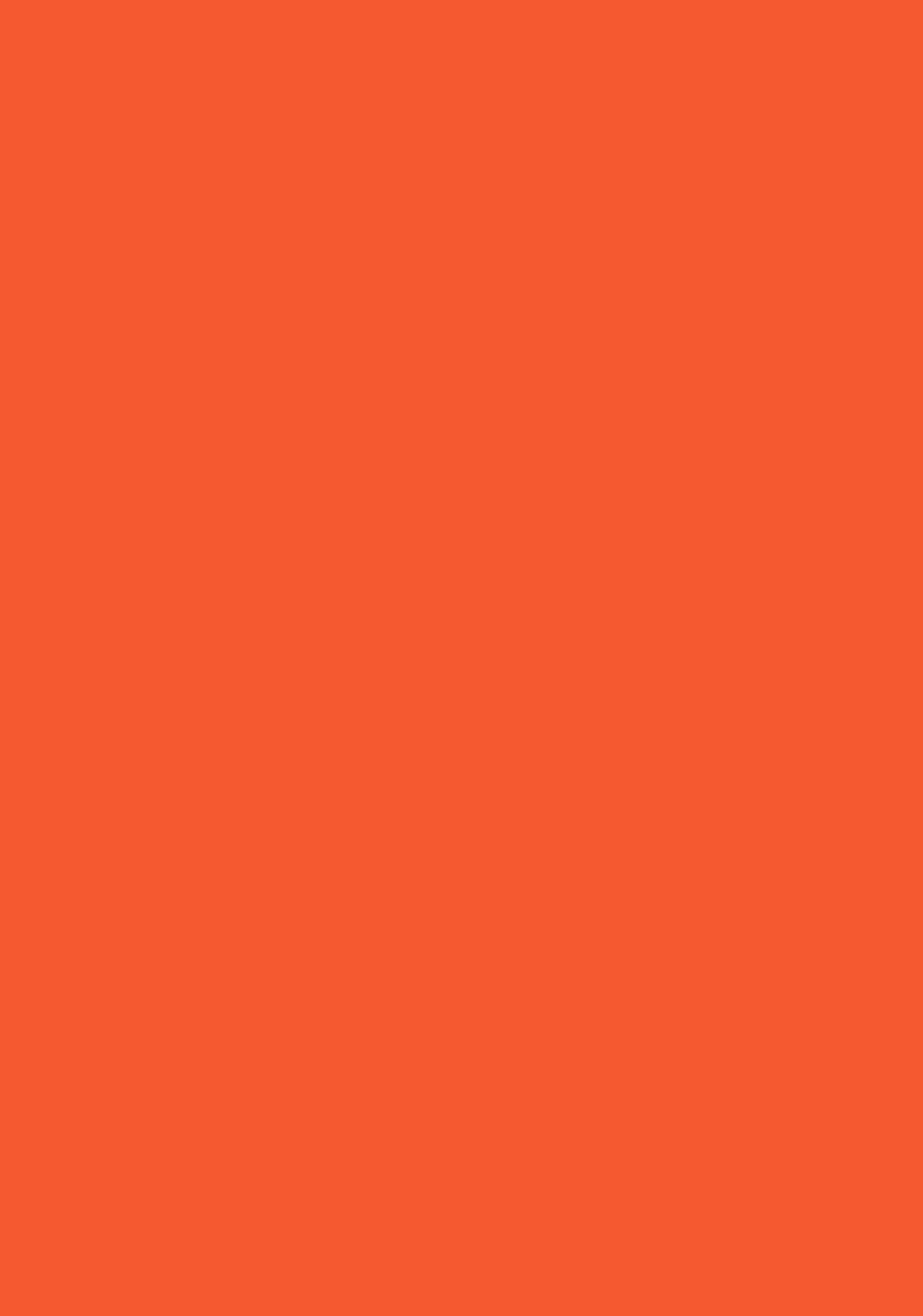